\documentstyle[prl,aps,epsf,floats,twocolumn]{revtex}

\begin{document}
\twocolumn[\hsize\textwidth\columnwidth\hsize\csname @twocolumnfalse\endcsname
\title{Tests for Gaussianity of the MAXIMA-1 CMB Map
}

\author{
  J.~H.~P.~Wu$^{1}$,
  A.~Balbi$^{2,3,4}$,
  J.~Borrill$^{3,5}$,
  P.~G.~Ferreira$^{6}$,
  S.~Hanany$^{3,7}$,
  A.~H.~Jaffe$^{1,3,8}$,
  A.~T.~Lee$^{3,4,9}$,
  B.~Rabii$^{3,9}$,
  P.~L.~Richards$^{3,9}$,
  G.~F.~Smoot$^{3,4,8,9}$,
  R.~Stompor$^{3,8,10}$,
  C.~D.~Winant$^{3,9}$
}

\address{
$^{1}$Dept. of Astronomy, University of California,
  Berkeley, CA94720-3411, USA\\
$^{2}${Dipartimento di Fisica, Universit\`a Tor Vergata, Roma,
Via della Ricerca Scientifica, I-00133, Roma, Italy}\\
$^{3}${Center for Particle Astrophysics, University of
  California, Berkeley, CA94720-7304, USA}\\
$^{4}${Lawrence Berkeley National Laboratory,
 Berkeley, CA94720, USA}\\
$^{5}${National Energy Research Scientific Computing Center,
  Lawrence Berkeley National Laboratory, Berkeley, CA94720, USA}\\
$^{6}${Astrophysics, University of Oxford, Oxford, OX1 3RH,
 UK}\\
$^{7}${School of Physics and Astronomy, University of
  Minnesota/Twin Cities, Minneapolis, MN55455, USA}\\
$^{8}${Space Sciences Laboratory, University of California,
  Berkeley, CA94720, USA}\\
$^{9}${Dept. of Physics, University of California,
  Berkeley, CA94720-7300, USA}\\
$^{10}${Copernicus Astronomical Center, Bartycka 18, 00-716 Warszawa, Poland}
}


\maketitle
\begin{abstract}

Gaussianity of the cosmological perturbations in the universe
is one of the key predictions of standard inflation,
but it is violated by other models of structure formation 
such as topological defects.
We present the first test of the Gaussianity of the cosmic microwave background (CMB)
on sub-degree angular scales,
where deviations from Gaussianity are most likely to occur.
We apply the methods of moments, cumulants, the Kolmogorov test, the $\chi^2$ test,
and Minkowski functionals 
in eigen, real, Wiener-filtered and signal-whitened spaces,
to the MAXIMA-1 CMB anisotropy data.
We conclude that the data, which probe angular scales between 10 arcminutes
and 5 degrees, are consistent with Gaussianity.

\end{abstract}
\vskip .2in
]

\noindent
{\bf A.\ Introduction}:
The anisotropy in the cosmic microwave background (CMB) is
arguably the cleanest cosmic signal that preserves the intrinsic
statistical properties of cosmological perturbations \cite{HuSugSil}.
The recent observations from MAXIMA-1 \cite{Hanany} and BOOMERANG \cite{B98} 
have unambiguously detected a sharp peak in the CMB power spectrum.
This has favored inflation \cite{Guth} as the dominant
mechanism for structure formation of the universe \cite{Bouchet} 
as opposed to other
candidates like topological defects \cite{VilShe}. 
Another key prediction of standard inflation is that 
the distribution of cosmic perturbations are Gaussian,
while
other cosmological models such as isocurvature inflation (e.g.\ \cite{Pee})
and
topological defects (e.g.\ \cite{cs_ng}) predict otherwise.
Thus tests for the Gaussianity of CMB anisotropy data can 
discriminate between cosmological models.
In addition, a Gaussian distribution is an important ingredient in the 
estimation algorithm of CMB power spectra \cite{BJK,madcap}, which 
has been used to produce the recent MAXIMA-1 and BOOMERANG
results \cite{Hanany,B98}.
Adding the fact that
the estimation of cosmological parameters  \cite{Balbi,Lange,Jaffe00}
relies on the estimated CMB power spectra,
it is important to verify the Gaussian distribution of the CMB.

Tests for the Gaussianity of CMB data have been
carried out by numerous authors, mainly using data from the  
Differential Microwave Radiometer (DMR) on
the COsmic Background Explorer (COBE) \cite{smoot92}.  
Several statistics were used including
moments, cumulants, Minkowski functionals
(which include genus, e.g.\ \cite{Minkowski,genus}), 
the three-point function (e.g.\
\cite{threepoint}), bispectrum (e.g.\ \cite{Heavens98,FMG}), wavelet
transform (e.g.\ \cite{PVF}), etc.  
All these tests showed that
the data were consistent with Gaussianity, 
except for a couple of results \cite{FMG,PVF} that may have
non-cosmological origins \cite{notNG}.

In addition to the usual limitations from foreground contamination
and instrumental noise,
the $7^\circ$ angular resolution of the DMR data
is not ideal for tests of Gaussianity.
Angular resolution is an issue 
because the size of the causal horizon at last scattering is about one degree.
Thus in a sky patch of super-degree size,
there are many uncorrelated perturbations due to causality.
As a result, the central limit theorem guarantees that 
the pre-last-scattering anisotropy
on super-degree scales will tend to be Gaussian.
Even if the post-last-scattering anisotropy is strong and non-Gaussian,
it can be obscured by the pre-last-scattering contribution,
which we expect to be Gaussian.
Therefore a CMB map with sub-degree resolution can
provide much more powerful tests for Gaussianity.
Park et al.~\cite{PPRT} recently used a genus test 
on the QMAP and Saskatoon data,
which have a resolution of about $1.5^\circ$,
to show that they were consistent with Gaussianity. 
In this paper 
we report results from a series of Gaussianity tests on the
MAXIMA-1 CMB map \cite{Hanany}, which provides anisotropy information
on angular scales between $10'$ and $5^\circ$.
To optimize both the resolution and the signal-to-noise ratio for these tests,
we shall use a map with 5972 square pixels of $8'$ each \cite{Stompor}.  
Using these high-quality data we probe for the first time 
the Gaussianity of CMB anisotropy on sub-degree scales.

A CMB sky is Gaussian
if it is a realization of a process that 
is only specified by the two-point correlation function of 
a given cosmological model.
However,
because we have only one realization of the CMB sky,
it is not possible to exhaustively test 
the statistical properties of the process that generated it.
Therefore, we employ a Frequentist approach,
testing the null hypothesis of Gaussianity 
for the inflationary model that best fits the data.

\noindent
{\bf B.\ Karhunen-Lo\`{e}ve transform}:
We consider the Karhunen-Lo\`{e}ve (K-L) transform,
sometimes called Principal Component Analysis
or the signal-to-noise eigenmode transformation \cite{KL,BJK},
because it enables us 
not only to transform the observed CMB map into uncorrelated eigenmodes
of known signal-to-noise ratios,
and further to implement Gaussianity tests on the uncorrelated modes.
For the CMB,
it is standard to model the data ${\bf d}\equiv d_i$ as a linear sum
of uncorrelated signal ${\bf s}\equiv s_i$ and noise ${\bf n}\equiv n_i$,
with the correlation matrix ${\bf C\equiv\langle d d^{\rm T} \rangle=S+N}$
where ${\bf S=\langle s s^{\rm T} \rangle}$ and ${\bf N=\langle n n^{\rm T} \rangle}$.
In the noise-whitened space ${\bf d}^{\rm (W)}={\bf N}^{-1/2}{\bf d}$,
all the eigenvalues of the noise matrix, ${\bf N}^{-1/2}{\bf N}{\bf
  N}^{-1/2}={\bf I}$, are simply unity.
Thus
the eigenvalues ${\bf e}^{S({\rm W})}$ 
of the noise-whitened signal matrix ${\bf N}^{-1/2}{\bf S}{\bf N}^{-1/2}$
represent the square of signal-to-noise ratios of 
each eigenmode.
The coefficients ${\bf b}^{\rm (W)}$ of the noise-whitened eigenmodes in a data set
can be obtained by transforming ${\bf d}^{\rm (W)}$ to the basis 
which diagonalizes ${\bf N}^{-1/2}{\bf C}{\bf N}^{-1/2}$.
These coefficients are normally called the K-L coefficients.

\begin{figure}[t]
  \centering
  \leavevmode\epsfxsize=8cm \epsfbox{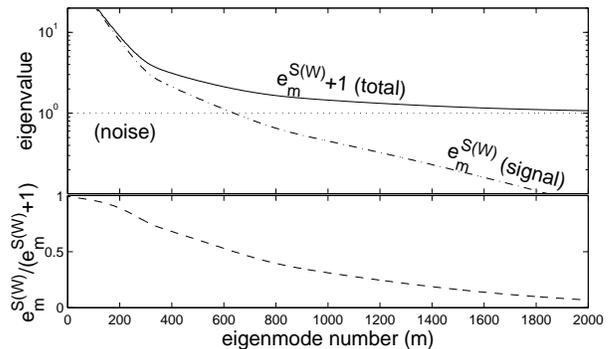}\\
  \caption
  []
  {Noise-whitened eigenvalues of the MAXIMA-1 CMB data (top),
   and the Wiener filter  (bottom).}
  \label{fig-eval}
\end{figure}

We compute ${\bf S}$ 
using the CMB power spectrum of
the best estimated cosmological model \cite{Balbi}
($\Omega_b=0.105$, $\Omega_c=0.595$, $\Omega_\Lambda=0.3$, and $h=0.53$)
and including the effects introduced by the beam shape
and the pixelization of the map \cite{asym-beam}.
The matrix ${\bf N}$ is estimated from the temporal data \cite{Stompor}.
The resulting eigenvalues in the noise-whitened space are shown 
in Fig.~\ref{fig-eval}, sorted in descending order. 
The dot-dashed line indicates
that only the first 639 modes
have signal-to-noise ratios $\left[{\bf e}^{S({\rm W})}\right]^{1/2}\geq 1$.
This number is well below our pixel number 5972.
It is determined by the signal and noise levels and the observing resolution
(beam) of a data set, but is independent of the pixel number of the map
when the pixel size is not larger than the observing resolution.

A common technique to remove the information of non-cosmological interest
in the map is Wiener filtering,
${\bf d}^{\rm WF}={\bf SC}^{-1}{\bf d}$.
This is equivalent to weighting the eigenmodes with the ratios
${\bf e}^{S({\rm W})}/({\bf e}^{S({\rm W})}+1)$ (see Fig.~\ref{fig-eval}).
The sum of these ratios for all the eigenmodes is 837,
again well below the total number 5972 of the eigenmodes.
We shall employ this technique in section D 
to amplify the statistical significance of the CMB signal in our map.

The K-L transform can also be used to test for Gaussianity.
If the underlying map is Gaussian,
then the eigenvalue-normalized K-L coefficients 
 ${\bf a}^{{\rm (W)}} \equiv 
{\bf b}^{{\rm (W)}}/({\bf e}^{S {\rm (W)}}+1)^{1/2}$ 
should be a set of Gaussian variables with mean zero and variance one.
We compute the ${\bf a}^{{\rm (W)}}$ of our data,
and use the observed frequency distribution to approximate
its one-point probability distribution functions (PDF) $p(\nu)$,
where $\nu$ is the number of standard deviations from the mean, for
both the entire 5972 and the first 639 modes
(i.e., those modes with signal dominating over noise).  
Both cases easily
pass the $\chi^2$ and the Kolmogorov tests for Gaussianity at
95\% confidence. 

\begin{figure}[t]
  \centering
  \leavevmode\epsfxsize=8cm \epsfbox{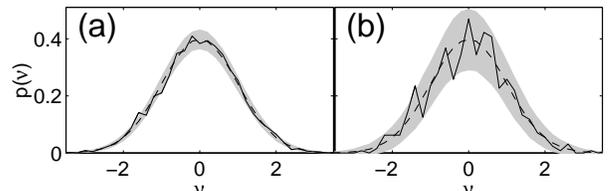}\\
  \caption
  []
  {PDF $p(\nu)$ of
   the entire 5972 (a) and the first 639 (b)
   ${\bf a}^{{\rm (W)}}$ of the MAXIMA-1 data (solid),
   the Gaussian expectation (dashed), and
   its 95\% CRG (shaded).
}
  \label{fig-pdf-e}
\end{figure}

In addition to the above tests,
we also use Monte Carlo (MC) simulation
to build a Gaussian reference frame
for the Frequentist approach.
We generate 100,000 Gaussian realizations, 
each of which is obtained by $d^{\rm MC}_i=C^{1/2}_{ij}g_j$,
where $g_j$ is a Gaussian variable of mean zero and variance one.
As a first application,
we use the MC simulation to find 
the probability distribution of $p(\nu)$ at each $\nu$
for the entire 5972 and the first 639 
 ${\bf a}^{{\rm (W)}}$.
Fig.~\ref{fig-pdf-e} shows that in both cases
the real data lie well within the 95\% confidence regions of Gaussianity
(hereafter CRG). 
We also compute the
moments and cumulants of the first 639 ${\bf a}^{{\rm (W)}}$
up to tenth order,
and they are all well within the 99\% CRG. 
All these results support the conclusions not only that
our map is consistent with Gaussianity,
but also that
our estimations of noise and CMB power spectrum 
(giving ${\bf N}$ and ${\bf S}$ respectively)
are consistent with the data so as to provide the proper eigenmodes
for the K-L transform.

\noindent
{\bf C.\ Minkowski functionals}:
The concept of Minkowski functionals is based on integral geometry.
According to Hadwiger's Theorem
$\beta+1$ Minkowski functionals are sufficient
to measure the morphology of a $\beta$-dimensional pattern.
In the case of CMB
  we have $\beta=2$ and thus 
  need only three Minkowski Functionals.
First we define
the excursion set $Q$ in a given CMB field as
the region in which the CMB amplitude $d_i$ is larger than a threshold:
$Q\equiv Q(\nu)=\{d_i| (d_i-\mu)/\sigma>\nu\}$,
where $\mu=\langle d \rangle$ and
$\sigma^2=\langle d^2 \rangle - \mu^2$.
Then the surface densities $v_i(\nu)\equiv V_i(Q)/A$
of Minkowski functionals $V_i(Q)$
for a CMB patch of angular area $A$ can be defined as
\begin{equation}
  \label{mf_cmb}
  v_0 = \frac{1}{A} \int_Q dA,\;
  v_1 = \frac{1}{4A} \int_{\partial Q} dl,\;
  v_2 = \frac{1}{2\pi A} \int_{\partial Q} \kappa dl,
\end{equation}
where 
$\partial Q$ is the boundary of the region $Q$,
$dA$ and $dl$ are the differential elements of $Q$ and $\partial Q$ 
respectively,
and $\kappa$ is the geodesic curvature of $dl$.
These Minkowski Functionals have different morphological meanings:
$V_0$ is the total area of $Q$, 
$V_1$ is the total length of its boundary, 
and 
$V_2$ is the number of isolated regions (hot spots)
in $Q$ minus the number of holes (cold spots).
For an isotropic Gaussian field
these functionals are characterized only by the field's variance $\sigma^2$
and the variance of its gradient $\tau=\langle |{\bf \nabla} d|^2 \rangle / 2$,
i.e.\
$v_{0\rm (G)}={\rm erfc}(\nu/\sqrt{2})/2$,
$v_{1\rm (G)}=\tau^{1/2}\exp(-\nu^2/2)/8\sigma$,
and $v_{2\rm (G)}=\tau \nu\exp(-\nu^2/2)/(2\pi)^{3/2}\sigma^2$.
We also note that the commonly used
one-point PDF $p(\nu)$
and
genus $g$,
or the `Euler-Poincar\'{e} characteristics',
are simply related to the Minkowski Functionals.
For the two-dimensional CMB,
  we have
  $p(\nu)=-\partial V_0 / A \partial \nu$ 
and
  $g=V_2+{V_0}/{2\pi}$. 

\begin{figure}[t]
  \centering
  \leavevmode\epsfxsize=6.5cm \epsfbox{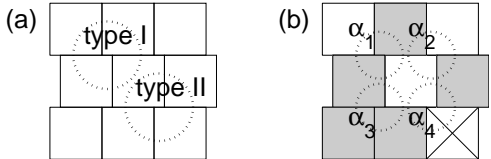}\\
  \caption
  []
  {(a) Vertices of type I and II formed 
    by inducing an offset in the pixel positions every other row.
   (b) An example of a pixelized CMB map with the artificial pixel offset.
}
  \label{fig-v2scheme}
\end{figure}

Because the MAXIMA-1 map has an irregular boundary 
and all pixels have a square shape lying on a regular lattice,
we need a numerical scheme to approximate its Minkowski Functionals.
For a given threshold $\nu$ and thus a given $Q$ on the lattice,
$v_0$ is simply the number of pixels in $Q$ 
divided by the total number of pixels in the map.
Here $v_1$ is determined by the length of $\partial Q$ multiplied by $\pi/4$
to correct for the square-pixel lattice effect.
To compute $v_2$,
we first induce offset in the pixel positions every other row.
As Fig.~\ref{fig-v2scheme}~(a) shows,
this gives two types (I and II) of vertices.
For each type,
the integration of $\kappa$ in equation (\ref{mf_cmb})
can be easily obtained by examining the status of the neighboring four pixels
(basically the integration is determined by
the change of the internal angle along $\partial Q$).
Fig.~\ref{fig-v2scheme}~(b) shows an example,
where the pixels in $Q$ are shaded and the one with no CMB data is labeled
with a cross.
In this case
the vertices $\alpha_1$, $\alpha_2$, $\alpha_3$ and $\alpha_4$
have unambiguously
$\int_{\alpha_i}(2\kappa/\pi) dl=-2$, $2$, $-1$ and $0$ respectively,
while 
a confusion in this calculation would occur
at $\alpha_1$and  $\alpha_2$
if there was no offset every other row.
We also notice that any bias possibly induced by the above algorithms
should disappear when the number of pixels is large,
as in our case.

\begin{figure}[t]
  \centering
  \leavevmode\epsfxsize=8.5cm \epsfbox{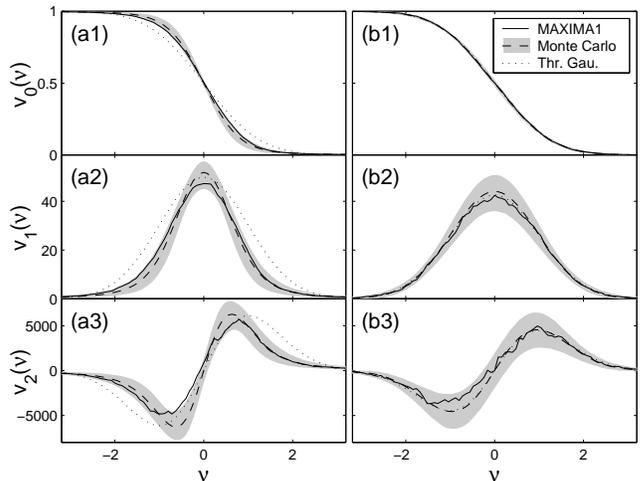}\\
  \caption
  []
  {Minkowski functionals (solid) of 
   the entire MAXIMA-1 map (a) and the central part (b).
   The Gaussian expectation obtained from the MC simulation (dashed),
   its $95\%$ CRG (shaded),
   and 
   the analytic Gaussian forms (dotted).
    }
  \label{fig-min0}
\end{figure}

We compute the $v_0(\nu)$, $v_1(\nu)$, and $v_2(\nu)$
for both the entire map
and the central $37\%$
(2209 pixels,  covering about $6.3^\circ\times 6.3^\circ$),
expected to have the lowest noise.
We use pixel sizes of $8'$, $16'$, and $24'$,
the last two obtained by averaging the neighboring pixels of the original map.
The results of all six cases lie within the $95\%$ CRG
obtained from the MC simulation described earlier. 
Fig.~\ref{fig-min0} shows the results of the two cases with $8'$ pixels.
We note that 
while the means of the MC simulation (dashed lines) are close to
the analytical isotropic Gaussian forms (dotted lines) in (b1--b3),
they deviate significantly from each other in (a1--a3).
This is due to the higher noise level near the edge of the map,
contributing as an anisotropic component in the map
(the RMS noise of the full map and the central part
are about $162\mu K$ and $57\mu K$ respectively).

We also note that
some results (solid lines) appear to have systematic departure from
the Gaussian expectations $\bar{v}^{\rm MC}_i(\nu)$ (dashed lines).
For example, (b3) shows the number of cold spots is less than 
the Gaussian expectation. 
To determine whether these discrepancies are really systematic,
we first define
\begin{equation}
  \label{I_i}
  I_i(\nu)=\int_{-\infty}^{\nu} 
  \frac{v_i(\nu')-\bar{v}^{\rm MC}_i(\nu')}
       {\sigma^{\rm MC}_i(\nu')}
  d\nu', \quad i=0,1,2,
\end{equation}
where $\sigma^{\rm MC}_i(\nu')$ is the standard deviation of $v_i(\nu')$
estimated from the MC simulation.
Applying this to both the real data and the MC simulation,
we obtain respectively the $\hat{I}_i(\nu)$ and 
its Gaussian-expected PDF $p(I_i(\nu))$.
The results from all six cases lie well within the 95\% CRG. 
Thus we know that
the apparent systematic deviations from Gaussianity in $v_i$
are statistically insignificant
as measured by $I_i(\nu)$.

\begin{figure}[t]
  \centering
  \leavevmode\epsfxsize=8.5cm \epsfbox{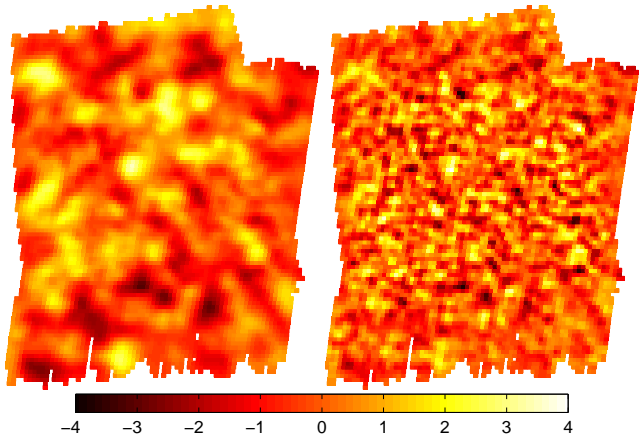}\\
  \caption
  []
  {The Wiener-filtered (left) and the signal-whitened (right)
   MAXIMA-1 map, with ranges of fluctuations of
   $(-3.1,3.4)\sigma$ and $(-3.7,4.3)\sigma$ respectively.
}
  \label{fig-map}
\end{figure}

\begin{figure}[t]
  \centering
  \leavevmode\epsfxsize=8.5cm \epsfbox{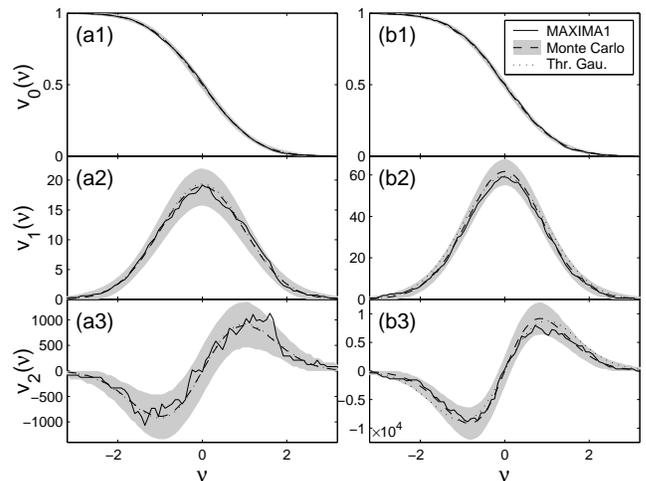}\\
  \caption
  []
  {Similar to Fig.~\ref{fig-min0}, but for the Wiener filtered (a) and
   the signal-whitened (b) MAXIMA-1 central map.
    }
  \label{fig-minw}
\end{figure}

\noindent
{\bf D.\ Wiener filtering and signal-whitening techniques}:
To increase the statistical significance of the CMB signal in the map,
we now consider two filtering methods.
One is the Wiener filtering addressed earlier,
and the other is a new signal-whitening technique 
${\bf d}^{\rm W}={\bf S}^{1/2}{\bf C}^{-1}{\bf d}$ \cite{whitenCMB},
which not only removes the anisotropy on scales where the noise dominates
(as in  Wiener filtering)
but also equalizes the anisotropy amplitudes 
on scales where the CMB signal dominates.
The signal-whitening procedure will reveal in ${\bf d}^{\rm W}$
the features of the non-Gaussian components
whose contribution in the CMB anisotropy dominates the Gaussian one
within at least a range of
the accessible scales \cite{whitenCMB}.
We apply these filtering methods to both the real map (Fig.~\ref{fig-map})
and the MC simulation,
and then compute their Minkowski functionals.
We find that for both the entire map and the central part (as previously)
the $v_i(\nu)$ and $I_i(\nu)$ of the filtered maps are within the 95\% CRG.
Fig.~\ref{fig-minw} shows results of the filtered central maps.
We also verify for both of the filtered maps that
none of the pixels with amplitude outside the $\pm 2\sigma$ range
coincides with the locations of any known radio or IRAS point sources \cite{roberta}.
Thus we have no statistically significant detection 
of localized non-Gaussianity in our data.



\noindent
{\bf E.\ Conclusion}:
We employ
moments, cumulants, the Kolmogorov test, the $\chi^2$ test,
and Minkowski functionals 
in eigen, real, Wiener-filtered and signal-whitened spaces
to implement a total of 82 (not independent) hypothesis tests for  Gaussianity
(22 in Sec.~B, 36 in Sec.~C, 24 in Sec.~D),
and show that the MAXIMA-1 CMB map is consistent with Gaussianity
on angular scales between $10'$ and $5^\circ$.
This gives us confidence in the Gaussian distribution used
in our estimation of the CMB angular power spectrum \cite{Hanany}, 
and consequently in the estimation of cosmological parameters \cite{Balbi,Jaffe00}.
Although our results are consistent with standard inflation
and
the CMB power spectrum of recent observations has favored inflation
as the dominant mechanism for structure formation \cite{Bouchet},
the existence of topological defects is still possible.
More sophisticated methods or
data with even higher resolution and signal to noise ratio
will be required to
fully explore this possibility. 

\noindent
{\bf {Acknowledgments}}---JHPW and AHJ acknowledge support from 
NASA LTSA Grant no.\ NAG5-6552 and NSF KDI Grant no.\ 9872979. 
PGF acknowledges support from the Royal Society. 
RS and SH acknowledge support from 
NASA Grant NAG5-3941. 
BR and CDW acknowledge support from NASA GSRP
Grants no.\ S00-GSRP-032 and S00-GSRP-031.
MAXIMA is supported by NASA Grant
NAG5-4454 and by the NSF through the Center for Particle
Astrophysics at UC Berkeley, NSF cooperative agreement AST-9120005.



\end{document}